\documentclass[aps]{revtex4}
\usepackage{graphicx}

\def\bra{\langle}
\def\ket{\rangle}
\def\<{\langle}
\def\>{\rangle}
\begin{document}
\title{Density Matrix in Quantum Mechanics and Distinctness of Ensembles Having
the Same Compressed Density Matrix}
\author{Gui Lu Long$^{1,2}$,
Yi-Fan Zhou$^1$, Jia-Qi Jin$^{1}$,  Yang Sun$^{3,1}$, Hai-Woong Lee$^4$}
\address{ $^{1}$Key Laboratory For Quantum Information and Measurements,  and
Department of Physics, Tsinghua University, Beijing,
100084, P R China\\
$^2$ Key Laboratory for Atomic and Molecular NanoSciences,
Tsinghua University, Beijing 100084, China\\
$^3$ Department of Physics, University of Notre Dame, Notre Dame,
IN 46556, USA\\
$^4$Department of Physics, KAIST, Dajeon 305-701, South Korea\\
Corresponding author: Gui Lu Long, Email: gllong@tsinghua.edu.cn
\mbox{$\;\;\;\;\;\;\;\;\;\;\;\;\;\;\;\;\;$}
\\
\mbox{$\;\;\;\;\;\;\;\;\;\;\;\;\;\;\;\;\;\;\;\;\;\;\;\;\;\;\;\;\;\;\;\;\;\;\;\;\;\;
\;\;\;\;\;\;\;\;\;\;\;\;\;\;\;\;\;\;\;\;\;\;\;\;\;\;\;\;\;\;\;\;\;\;\;\;\;\;\;\;\;\;\;\;
$}\\
\mbox{$\;\;\;\;\;\;\;\;\;\;\;\;\;\;\;\;\;\;\;\;\;\;\;\;\;\;\;\;\;\;\;\;\;\;\;\;\;\;
\;\;\;\;\;\;\;\;\;\;\;\;\;\;\;\;\;\;\;\;\;\;\;\;\;\;\;\;\;\;\;\;\;\;\;\;\;\;\;\;\;\;\;\;
$}\\
\mbox{$\;\;\;\;\;\;\;\;\;\;\;\;\;\;\;\;\;\;\;\;\;\;\;\;\;\;\;\;\;\;\;\;\;\;\;\;\;\;
\;\;\;\;\;\;\;\;\;\;\;\;\;\;\;\;\;\;\;\;\;\;\;\;\;\;\;\;\;\;\;\;\;\;\;\;\;\;\;\;\;\;\;\;
$}
}

\begin{abstract}
 We clarify
different definitions of the density matrix by proposing the use
of different names, the full density matrix for a single-closed
quantum system, the compressed density matrix for the averaged
single molecule state from an ensemble of molecules, and the
reduced density matrix for
 a part of an entangled quantum system, respectively. We show that
 ensembles with the same compressed density matrix can be
physically distinguished by observing fluctuations of various observables. This is in
contrast to a general belief that ensembles with the same compressed density matrix are
identical. Explicit expression for the fluctuation of an observable in a specified
ensemble is given.  We have discussed the nature of nuclear magnetic resonance quantum
computing.  We show that the conclusion that there is no quantum entanglement in the
current nuclear magnetic resonance quantum computing experiment is based on the
unjustified belief that ensembles having the same compressed density matrix are
identical physically.  Related issues in quantum communication are also discussed.
\mbox{$\;\;\;\;\;\;\;\;\;\;\;\;\;\;\;\;\;\;\;\;\;\;\;\;\;\;\;\;\;\;\;\;\;\;\;\;\;\;
\;\;\;\;\;\;\;\;\;\;\;\;\;\;\;\;\;\;\;\;\;\;\;\;\;\;\;\;\;\;\;\;\;\;\;\;\;\;\;\;\;\;\;\;
$}\\
\mbox{$\;\;\;\;\;\;\;\;\;\;\;\;\;\;\;\;\;\;\;\;\;\;\;\;\;\;\;\;\;\;\;\;\;\;\;\;\;\;
\;\;\;\;\;\;\;\;\;\;\;\;\;\;\;\;\;\;\;\;\;\;\;\;\;\;\;\;\;\;\;\;\;\;\;\;\;\;\;\;\;\;\;\;
$}\\
\mbox{Keywords: full density matrix, compressed density matrix, reduced density matrix,
distinction of ensembles}
\end{abstract}
 \maketitle

\begin{enumerate}

\section*{1. INTRODUCTION}

\item\label{ss1} In a quantum system, the wave function specifies
all the physical properties of the system. The wave function
provides the maximal information. A system with a wave function
description is said to be in a pure state. In contrast sometimes
there is incomplete information about a quantum system and this
may be more common. Von Neumann and Landau proposed to use the
density matrix to describe the state of a quantum system in this
situation\cite{von1,landau}. It has been widely used nowadays, and
is presented in standard textbooks and monographs on quantum
mechanics, for instance in Refs.
\cite{preskill,peres,despagnat,fano,ka}. With the development of
quantum information in recent years, there have been renewed
interests in the density matrix formalism. It is the basic tool
for the study of quantum entanglement, channel capacity in quantum
communication and so on.

\item\label{ss3} Though the density matrix has been proposed for
nearly 80 years, there are still confusions and sometime disputes
on matters related to it. Some are caused by a confusion in the
terminology, but others are true disputes of the fundamental
nature.  A notable example is the question of distinctness of
ensembles having the same density matrix. These matters are
becoming more acute with the progress in quantum computation
research and  to cause true concerns. Whether current liquid
nuclear magnetic resonance (NMR) quantum computation is genuine
quantum mechanical or just a classical simulation is just one such
problem.

\item\label{ss4} In this article,we shall carefully inspect the definition of the
density matrix, and that of the mixed state. This will remove a large number of
disagreements. We propose to use three different terminologies for the density matrix
in different situations: the full density matrix to describe the state of a single,
closed  quantum system, it always describes  a pure state and is equivalent to the wave
function; a reduced density matrix for the state of part of a coupled quantum system,
it usually describes a mixed state; a compressed density matrix for an ensemble of
independent quantum systems. After the classification of the density matrix, we will
examine the distinctness of ensembles having the same compressed  density matrix. We
will show that ensembles having the same compressed density matrix can be distinguished
physically. We show that this does not contradict one of the fundamental postulates of
quantum mechanics that the density matrix completely specifies all the properties of a
quantum ensemble, as stated by Asher Peres\cite{peres1}. We then apply this result to
the question of the nature of current NMR quantum computation, and show that the
conclusion that there is no entanglement in the current NMR quantum computing is based
on the unjustified belief that ensembles having the same compressed density matrix are
physically identical, and hence is invalid. The implication of this result in quantum
key distribution and other areas of quantum information are also discussed.

\section*{2. DENSITY MATRIX AND MIXED STATE}
\label{s2}

\item First we introduce some notations in order to avoid
confusion because different people use different definitions.

{\bf A molecule} in this article refers to a single quantum system
such as a qubit, a composite quantum system and so on.

{\bf An ensemble} in this article is defined as a collection of
$N$ non-interacting molecules of the same kind in the same or
different quantum states. An ensemble contains $N$ non-interacting
molecules, and there are $N_i$ number of molecules in state
$|\psi_i\ket$, $i=1,\cdots,k$, $\sum_{i=1}^kN_i=N$.

This definition of an ensemble has been used by Shankar,
Merzbacher, d'Espagnat, Von
Neumann\cite{shankar,merzenbacher,despagnat,von2}.  A molecule is
a quantum system which may contain a complex internal structure,
for instance an atom, a molecule, a Bose-Einstein condensate all
may be viewed as a molecule in an ensemble. The molecules are of
the same kind, but they are independent of each other and can be
distinguished, for example by their position, hence they are not
identical particles. Usually they are far apart, so that their de
Broglie waves rarely overlap.

We note that the definition of ensemble by Asher Peres is
different from ours \cite{peres}. In his definition, an ensemble
is defined by a probability distribution and there are infinite
number of molecules in an ensemble.

It is important to perceive the differences in the definition of ensemble in
statistical physics and quantum mechanics. In statistical mechanics \cite{pathria}, an
ensemble is a collection of infinite number of the system under identical physical
constraints. For instance, a micro-canonical ensemble is specified by the particle
number $N$ and the total energy $E$. Then a system with $N_i$ number of molecules in
state $|\psi_i\ket$, and $N=N_1+N_2+\cdots+N_i+\cdots $  satisfying the given
constraints on the particle number and total energy will be a member in the statistical
ensemble. Whereas in quantum mechanics, the system of $N$ molecules itself is now an
ensemble, each molecule is treated as a system and the density matrix $\rho$ describes
the state of averaged molecule in this system. This different comprehension of
ensembles in statistical mechanics and quantum mechanics causes many of the
misunderstanding.

{\bf A coupled  quantum system} is a quantum system with several
constituents. For instance, a molecule is a coupled quantum
system, and each atom in a multi-atom molecule is a constituent.
The constituents in a coupled quantum system may be confined, for
instance the atoms in a molecule,  or far away apart, for instance
two entangled photon pairs forming a coupled quantum system.

{\bf By a sampling measurement of observable $\Omega$}, we mean the following. We
select randomly one molecule from an ensemble and makes an $\Omega$ measurement on this
molecule. Then the ensemble is restored to its original state, namely $N_i$ molecules
in state $|\psi_i\ket$, and we randomly select a molecule again and makes an $\Omega$
measurement on the molecule. We repeat this process sufficient number of times. Then
the result of the sampling measurement of $\Omega$ is the averaged value of these
individual measurements, and it is denoted by $\bra \Omega \ket$. In the density matrix
formalism, the expectation value of an observable $\Omega$ in an ensemble,
$\bra\Omega\ket={\rm Tr}(\rho\Omega)$, should be understood in this way.

In contrast, {\bf by a global measurement of observable $\Omega$},
we mean a measurement on the whole ensemble: we fetch every
molecule in the ensemble and measure its $\Omega$.  The result of
the global measurement of $\Omega$ will be the sum of all these
individual measurements. The fluctuation of the global measurement
of $\Omega$ is understood in the same way as that for a quantum
system: one prepares many copies of the same ensemble, he makes
global measurement of $\Omega$ on each of these ensembles. The
expectation value of the global measurement will be the averaged
value of these global measurements, and the fluctuation is the
standard deviation from this averaged (or mean) global
measurement. We denote the global measurement by a subscript $E$.
The expectation value of a global measurement of $\Omega$ is
denoted by $\bra \Omega \ket_E$. For instance, for an ensemble of
nuclear spins, the $z$-component of the spin operator $\Sigma_z$
for the whole ensemble is the sum of all the individual nuclear
spins,
$$\Sigma_z=\sum_{i=1}^N \sigma_z(i).$$
One can obtain its value by measuring the total magnetic polarization of the whole
ensemble, thus sometimes one can get the global measurement result without measuring
each of the individual molecule.

The expectation value of a global measurement and the sampling
measurement is related by the following
\begin{eqnarray}
\bra \Omega \ket_E=N\bra \Omega \ket,
\end{eqnarray}
where $N$ is the number of molecules in the ensemble.

\item\label{ss5} Von Neumann defined a density matrix and mixed
and pure states in the following way\cite{von2}.  Consider the
case in which one does not know what state is actually present in
a system, for example when several states $|\phi_1\ket$,
$|\phi_2\ket$, $\cdots$ with respective probabilities $w_1$,
$w_2$, $\cdots$, ($w_1\ge 0$, $w_2\ge0$, $\cdots$,
$w_1+w_2+\cdots=1$) constitute the description of the system. Then
the expectation value of a physical observable $\Omega$ in the
system is
\begin{eqnarray}
\bra \Omega\ket=\sum_n w_n\bra\phi_n|\Omega|\phi_n\ket={\rm
Tr}(\rho\Omega).
\end{eqnarray}
The density operator, or density matrix
\begin{eqnarray}
\rho=\sum_nw_n|\phi_n\ket\bra\phi_n|,
\end{eqnarray}
is used to describe the state of the quantum system under study.
It characterizes the mixture of states just described completely
{\bf with respect to its statistical properties}.

When  only one $w_i$ is equal to one and the rest $w_i$'s are
zero, then the quantum system is described by a single wave
function, and the system is said to be in a pure state. When  more
than one $w_i$'s are nonzero, the state is called a mixed state.

The above notion is commonly accepted. However, in practice there
are different scenarios where the density matrix can be applied.
For example, the abstract definition of density matrix is applied
to an ensemble with $N$ molecules. If in an ensemble of $N$
independent molecules, there are $N_i$  molecules in state
$|\psi_i\ket$, then the ensemble is described by the following
density matrix
\begin{eqnarray}
\rho=\sum_{i=1}^m w_i|\psi_i\ket\bra \psi_i|,
\end{eqnarray}
where $w_i=N_i/N$ and $m$ is the number of possible single
molecule wave functions in the ensemble.

Here the density matrix represents the "state" of an averaged molecule from this
ensemble: a molecule in this ensemble has $w_i=N_i/N$ probability in state
$|\psi_i\ket$. $w_i$ is the probability a molecule is in state $|\psi_i\ket$ if one
fetches a molecule from the ensemble randomly. If $m>1$, the ensemble is said to be a
mixed ensemble, or simply a mixed state. If $m=1$, the ensemble is an ensemble in which
all molecules are in the same state, and the ensemble is called a pure ensemble, or
simply a pure state.

This definition of mixed state is widely adopted in textbooks and monographs of quantum
mechanics, for example in Shankar\cite{shankar}, Merzbacher \cite{merzenbacher} and so
on. d'Espagnat has suggested to use the term, proper mixture for this type of
object\cite{despagnat}, and this practice is adopted in this article.

This definition of mixed state is also used by Asher Peres
\cite{peres} and Preskill\cite{preskill}. For example, in one of
Peres's example, there are two different ensembles that have the
same density matrix $\rho={1\over 2}{\rm \bf 1_2}$, where ${\rm
I_2}$ is a 2 by 2 unit matrix. One ensemble is prepared by
preparing each qubit in either state $|0\ket$ or $|1\ket$
according to the result of a coin-tossing. The other ensemble is
prepared by making the state of each qubit in either
\begin{eqnarray}
|\pm\ket=(|0\ket\pm|1\ket)/2, \end{eqnarray}
 also according to the
result of a coin-tossing.  Though each individual qubit from both
ensembles is in a definite quantum state, but {\bf on average}, a
qubit from the first ensemble has half probability in state
$|0\ket$ and $|1\ket$  respectively, namely if one fetches a qubit
from the prepared ensemble, there is half probability the qubit is
in state $|0\ket$ and another half probability in state $|1\ket$.
Similar statement could also be made for the second ensemble. One
should note that in Peres's example, the ensemble is defined to
have infinite number of molecules. In this article, we talk of
ensembles with a finite number of molecules.

\item\label{ss7} Landau proposed the density matrix in a different
way \cite{landau}. In a paper on damping problems in quantum
mechanics, Landau introduced the density matrix to study coupled
quantum systems. Suppose there are two systems, A and B,  coupled
together. The wave function for the coupled AB-system is
\begin{eqnarray}
|\psi_{AB}\ket=\sum_{n,r}c_{nr}|\psi_n\ket|\psi'_r\ket.
\end{eqnarray}

The expectation value of an observable ${\Omega}$ pertinent to the
A-system in this state is
\begin{eqnarray}
\bra {\Omega^A}\ket=\bra
\psi_{AB}|{\Omega^A}|\psi_{AB}\ket=\sum_{n,m}\rho_{A,nm}\Omega^A_{nm}={\rm Tr}(\rho_A
\Omega^A),
\end{eqnarray}
where
\begin{eqnarray}
\Omega^A_{nm}=\bra \psi_n|\Omega^A|\psi_m\ket,
\end{eqnarray}
and
\begin{eqnarray}
\rho_{A,nm}=\sum_{r} c^{\ast}_{nr}c_{mr}={\rm
Tr}_B(|\psi_{AB}\ket\bra\psi_{AB}|).
\end{eqnarray}
The density matrix $\rho_A$ describes the "state" of the A-system. When the rank of the
density  matrix is greater than 1, the state is also a mixed state.  This type of mixed
state is different from the proper mixed state, although they have the same
mathematical structure. d'Espagnat suggested to use the terminology, improper mixture,
for this type of object. We also adopt this  in this article.

\item There are two different situations that a probabilistic
description is required. In one case, the description is due to
our lack of information about the quantum system, but the quantum
system itself is actually in a definite quantum state described by
a wave function. For instance in quantum key distribution
\cite{bb84}, Alice prepares a sequence of qubits, each qubit is
randomly prepared in one of the following states: $|0\ket$,
$|1\ket$, $|+\ket$, $|-\ket$. To anyone other than Alice, each
photon is in a proper mixed state described by $\rho={1\over
2}{\rm\bf I_2}$. But to Alice, it is in a definite pure state
because she has prepared it herself.

In another situation, a probabilistic description is the best one can provide. Suppose
that  two qubits $A$ and $B$ are in a Bell-basis state,
\begin{eqnarray}
|\phi^{+}_{AB}\ket=\sqrt{1\over 2}\left\{|0_A 0_B\ket+|1_A
1_B\ket\right\}.
\end{eqnarray}
If one wants to describe the state of qubit $A$ without mentioning
qubit $B$, the best one can do is to use the density matrix
$\rho_A={1\over 2}{\rm\bf I_2}$. If one measures the state of
qubit there is half probability each to obtain $|0\ket$ or
$|1\ket$. Anyone, without exception, will get this result. This is
in contrast to the previous example in which Alice has full
knowledge of each qubit.

\item There is an essential difference between the proper mixed
state and the improper mixed state. In the proper mixed state,
each individual molecule in an ensemble is in a definite quantum
state. In an improper mixed state, the molecule is actually in
such a state with the proposed probability distribution. In the
example when qubits A and B  are in a Bell-basis state,
$|\phi^{+}_{AB}\ket=(|0_A 0_B\ket+|1_A 1_B\ket)/\sqrt{2}$, the
reduced density matrix for qubit A is $\rho_A={1\over 2}{\rm\bf
1_2}$. In this state the qubit A is actually in state $|0\ket$ and
$|1\ket$ {\bf simultaneously}.

\item\label{ss2} The mathematical formalism of the density matrix
are well-known\cite{fano}. Its equation of motion in the
Schoedinger picture obeys the Liouville equation
\begin{eqnarray}
i\hbar{\partial \rho \over \partial t}=\left[H,\rho\right].
\end{eqnarray}
The density matrix $\rho$ is positively definite. Its properties
are summarized as follows
\begin{eqnarray}
{\rm Tr}(\rho)&=&1,\\
\rho^{\dagger}&=&\rho,\\
{\rm Tr}(\rho^2)&\le&1.
\end{eqnarray}
It should be pointed out that wave function description can also
be equivalently expressed in terms of a density matrix,
\begin{eqnarray}
\rho=|\Psi\ket\bra\Psi|.
\end{eqnarray}

The density matrix in this case of a pure state is characterized
by the equality  ${\rm Tr}(\rho^2)=1$. The density matrix with
${\rm Tr}(\rho^2)<1$ is a mixed state, either proper or improper.

\section*{3. THREE DIFFERENT DENSITY MATRICES AND MIXED STATES}
\label{s3}

\subsection*{3.1 The FULL DENSITY MATRIX for a single QUANTUM SYSTEM}

\item\label{ss8} The state of a single, closed quantum system is
always described by a state vector $|\psi\ket$. This is dictated
by the basic postulate of quantum mechanics \cite{shankar}: "the
state of the particle is represented by a vector $|\psi(t)\ket$ in
a Hilbert space." The density matrix for such a system is
$\rho=|\psi\ket\bra\psi|$. In this case, the density matrix
description and the wave function description are strictly
equivalent, and the unphysical overall phase in the wave function
description is naturally eliminated in the density matrix
description.  It contains {\bf ALL} the physical information about
the system. We call the density matrix in this case  a {\bf full
density matrix}.

A single, closed quantum system may have complicated structures.
For instance a Bose-Einstein condensate in an isolated environment
is considered as a single closed quantum system. An isolated atom
with many electrons and nucleons is another example. A quantum
system of $N$ 2-level molecules(or qubits)  could also be viewed
as a single quantum system if we want to describe all the details
of the whole $N$ qubits. The density matrix for such an $N$-qubits
system is of $2^N\times 2^N$ dimension.

Though the state of such a system is always pure, a mixed state
description for such a system is  possible when one considers only
the average properties of the system over a period of time, or
over some parameters. For instance if the system is under rapid
change in time and only average property of the system over a
relatively long period of time is concerned, the state of the
system could be described by a density matrix obtained by
averaging over the time
\begin{eqnarray}
\bar{\rho}=\int_{t_1}^{t_2}\lambda(t)\rho(t)dt,
\end{eqnarray}
where $\lambda(t)$ is some probability distribution function. If
the resulting density matrix describes a mixed state, it is a
proper mixed state because we have actually traded the space
averaging in an ensemble with the time averaging for a single
molecule over some time.

\subsection*{3.2 The COMPRESSED DENSITY MATRIX for an ENSEMBLE}

\item\label{ss9} We call the single-molecule density matrix  over an ensemble of $N$
independent molecules as a {\bf compressed density matrix}. If the compressed density
matrix is pure, it indicates that all the single molecules in the ensemble are in the
same quantum state. If the compressed density matrix is a mixed one, it represents an
ensemble with molecules in distinct quantum states. As stressed earlier, each
individual molecule in the ensemble is in a definite quantum state, rather than in
different states with respective probabilities. A mixed state arising from the average
over an ensemble is called a proper mixture by d'Espagnat\cite{despagnat} as we
mentioned earlier. Mixed state from this average is called the ignorance interpretation
by Cohen \cite{cohen}.

\subsection*{3.3 REDUCED DENSITY MATRIX in composite QUANTUM SYSTEM}

\item The density matrix obtained by tracing out partial degrees
of freedom of a coupled system is called {\bf  reduced density
matrix}. Suppose there are $M$ constituent molecules in a coupled
quantum system. The wave function of the coupled system is
$|\Psi_{A_1A_2...A_M}\ket$. The reduced density matrix for
molecule $A_i$ can be obtained by tracing out the degrees of other
molecules
\begin{eqnarray}
\rho_{A_i}={\rm Tr}_{A_1
A_2...A_{i-1}A_{i+1}...A_M}(|\Psi_{A_1A_2...A_M}\ket\bra\Psi_{A_1A_2...A_M}|).
\end{eqnarray}

One can also define reduced density matrix for two or more
molecules by tracing the degrees of freedom of the remaining
molecules, for instance, the reduced density matrix for molecules
$A_i$ and $A_j$ is defined as
\begin{eqnarray}
\rho_{A_iA_j}={\rm Tr}_{A_1
A_2...A_{i-1}A_{i+1}...A_{j-1}A_{j+1}...A_M}(|\Psi_{A_1A_2...A_M}\ket\bra\Psi_{A_1A_2...A_M}|).
\end{eqnarray}

The reduced density matrix has also been defined for identical
particle systems. For an $M$-identical-particle system, one can
define one-particle, two-particle, ... reduced density matrices as
follows \cite{cnyang}
\begin{eqnarray}
\rho^{(1)}&=&{\rm Tr}_{i_2i_3...i_M}(\rho),\nonumber\\
\rho^{(2)}&=&{\rm Tr}_{i_3...i_M}(\rho),\nonumber\\
& &\vdots\nonumber\\
 \rho^{(M-1)}&=&{\rm Tr}_{i_M}(\rho),
\end{eqnarray}
where $\rho$ is the density matrix for the coupled $M$-particle
system. The whole $M$-particle system is treated as a single
quantum system. The subscript $i_k$ is the single-particle state
index. These definitions are useful in the study of quantum
entanglement in identical particle systems\cite{schlieman,rlong}.

It is worth noting that an $M$-particle coupled quantum system is different from an
ensemble of $M$ independent molecules. Usually the particles in an $M$-particle coupled
quantum system are interacting and correlated. However the molecules in an ensemble
described by a compressed density matrix are independent and non-interacting. The
coupled $M$ particle system should be described by a $M$-particle density matrix, while
an ensemble of $M$ molecules is described by a single molecule compressed density
matrix. For example, a $M$-qubit system is described by a $2^M\times 2^M$ density
matrix, and an ensemble of $M$ qubits, such as $M$ spin-1/2 molecules is described by a
$2\times 2$ density matrix. In an NMR ensemble of  7-qubit molecules, there are usually
$N=O(10^{16})$ molecules. However the compressed density matrix for the ensemble is
$2^7\times 2^7$, rather than $2^{7N}\times 2^{7N}$ \cite{jcplong}.

An improper mixed state usually could not be identified with a
proper mixed state with the same density matrix\cite{despagnat},
as we see they represent different objects though their
mathematical descriptions are the same. This type of mixed state
was interpreted by Cohen as the ancilla interpretation
\cite{cohen}.

\item\label{ss11}
 It is useful to make a comparison between a single qubit and a
 coupled 2-qubit system to demonstrate the difference between a pure quantum
 state and an improper mixed state.
 Suppose a single qubit $A$ is in state
 \begin{eqnarray}
 |+x_A\ket=\sqrt{1\over 2}\left\{|0_A\ket+|1_A\ket\right\}.\label{e11.1}
 \end{eqnarray}
 Consider another system in which  qubit A and qubit B form an entangled state
 \begin{eqnarray}
 |\phi^{+}_{AB}\ket=\sqrt{1\over
 2}\left\{|0_A0_B\ket+|1_A1_B\ket\right\}.\label{e11.2}
\end{eqnarray}
If one measures $\sigma_{z}$ on qubit A, he will obtain $|0_A\ket$
and $|1_A\ket$ with 50\% probability in each case. But if he
measures $\sigma_{x}$ on qubit A, he will obtain with certainty
$|+x_A\ket$ in the first case as it is an measurement on an
eigenstate, whereas he will still have 50\% probability to obtain
$|+x_A\ket$ and $|-x_A\ket$ respectively in the latter case.

\section*{4. THREE SIMPLE EXAMPLES}
\label{s4p}

 \item \label{ss12} We study three  ensembles of qubits, that will be all described
 by the same mathematical density matrix $\rho={1\over 2}{\rm I_2}$ when $N$ approaches
 infinity,
  to illustrate
 the different mixed states and their relationship. Suppose there
 are $N$ pairs of qubits in state given by Eq. (\ref{e11.2}). All the $A$ qubits
 are given to Alice and all $B$ qubits to Bob. They form ensembles
 $A_1$ and $B_1$ respectively. Hence each $A$ qubit molecule in $A_1$ is
 in an improper mixed state described by reduced density matrix $\rho_{A}={1\over 2}I_2$, and
 similarly each $B$ qubit molecule in ensemble $B_1$ is also in an improper
 mixed state with the same reduced density matrix. It should be
 pointed out that all the $A$ qubit molecules in ensemble $A_1$ are
 in
 the same state, there is no difference between the states of the different molecules
  in ensemble $A_1$. The same conclusion
 can also be said about ensemble $B_1$. If one looks at the
 compositions of the ensembles, one could say that the ensembles
 are "pure", as they are made of molecules all in the same "state", though the
 "state" is not a pure quantum state described by a single state
 vector. This is a generalization of the usual pure state in which
 all the molecules are in the same quantum state described by a
 wave function.

 In the second example, one performs $\sigma_z$  measurement on
 each qubit in ensemble $B_1$. Because of the collapse of state in
 quantum mechanics, each qubit in $B_1$ collapses into either
 $|0\ket$ or $|1\ket$, with equal probabilities. Consequently, the
 corresponding $A$ molecules in ensemble $A_1$ also collapses into
 state $|1\ket$ or $|0\ket$ respectively. The ensembles after the
 measurement are called $A_2$ and $B_2$ respectively. In ensemble
 $B_2$, each qubit is in a definite quantum state and  the
 preparer of the ensemble who performs the $\sigma_z$ measurement knows all these information.
 He also knows the state of $A$
 molecules in ensemble $A_2$ as he can infer from the prior
 entanglement between qubit $A$ and $B$ and the
 result of his measurement of qubit B. Suppose after the measurement, the number of qubits in state
 $|0\ket$(or $|1\ket$) is $N/2-N_{\delta}$(or $N/2+N_{\delta}$), then the compressed density matrix
 of $A_2$ is
 \begin{eqnarray}
 \rho_{A_2}=\left(\begin{array}{cc} {1\over 2}-{N_\delta\over N} &
  0\\
                                      0                         &  {1\over 2}+{N_\delta\over
                                      N}\end{array}\right).\label{rhoa2}
                                      \end{eqnarray}
When $N$ approaches infinity, the
 compressed density matrices for qubit $A_2$ and $B_2$   all
become $\rho={1\over 2}{\rm I_2}$. $N_\delta$ is a random number,
can be  positive and negative or zero, and  is proportional to
$\sqrt{N}$ according to statistics.

 In the third example, one performs $\sigma_x$  measurement on
 each qubit in ensemble $B_1$. Again due to the collapse of state in
 quantum mechanics, each qubit in $B_1$ collapses into either
 $|+x\ket$ or $|-x\ket$, with equal probabilities. Consequently, the
 corresponding $B$ molecules in ensemble $A_1$ also collapses into
 state $|-x\ket$ or $|+x\ket$ respectively. The ensembles after the
 measurement are called $A_3$ and $B_3$ respectively. In ensemble
 $B_3$, each qubit is also in a definite quantum state and  the
 preparer of the ensembles who has performed the $\sigma_x$ measurement
 knows all these information. He also knows the state of $A$
 molecules in ensemble $A_3$. Suppose after the measurement, the number of qubit in
 $|+x\ket$ (or $|-x\ket$) is
 $N/2-N_x$ ( or $N/2+N_x)$, then the compressed density matrix of $A_3$ is
\begin{eqnarray}
\rho_{A_3}&=&\left({1\over 2}-{N_x\over N}\right)|+x\ket\bra +x|
+\left({1\over 2}+{N_x\over N}\right)|-x\ket\bra -x|\nonumber\\
&=& \left(\begin{array}{cc} {1\over 2} & -{N_x\over N}\\
                                       -{N_x\over N}          &  {1\over 2}\end{array}\right).
                                      \end{eqnarray}
 Again the
 compressed density matrices for qubit $A_3$ and $B_3$  are all
 approaching
$\rho={1\over 2}{\rm I_2}$ when $N$ approaches infinity. Here
$N_x$ is also random, could be positive, negative or zero, and is
proportional to $\sqrt{N}$.

We stress here that when the number of qubit $N$ is finite, the three ensembles are
different and they are distinguishable physically. When a measurement is performed on
an ensemble, the compressed density matrix of the resulting ensemble is usually
uncontrollable, and this is especially so when the number $N$ is small. Even if one
performs the measurement on the same observable, the resulting compressed density
matrix is different at different times. Suppose one has $N$ EPR pairs in state in Eq.
(\ref{e11.2}). As usual, the $A$ particles are given to Alice and the $B$ particles
given to Bob. When Bob performs $\sigma_z$ measurement on each of his $B$ particles, he
will get an ensemble with a compressed density matrix $\rho_B$. Suppose he repeats the
process a second time starting from state in Eq.(\ref{e11.2}), he will obtain another
ensemble with a compressed density matrix $\rho_B'$. Usually $\rho_B\neq \rho_B'$ as
can be seen in Eq.(\ref{rhoa2}). An extreme case is an ensemble $A_1$ with $N=2$.
Before any measurement, the state of any of the two particles is an improper mixed
state with density matrix, $\rho={1/2}{\rm I_2}$. Suppose $\sigma_z$ is measured on
each of the two particles. Then four different results may occur: 1) both qubits have
spins up, and the resulting compressed density matrix is a pure state,
\begin{eqnarray}
\left(\begin{array}{cc} 1 & 0\\
                        0 & 0\end{array}\right);
                        \end{eqnarray}
2) both qubits have spins down, and this gives a compressed
density matrix,
\begin{eqnarray}
\left(\begin{array}{cc} 0 & 0\\
                        0 & 1\end{array}\right);\end{eqnarray}
3) one qubit has spin up and the other spin down, and vice versa.
In both cases, the compressed density matrices of the resulting
ensemble are
\begin{eqnarray}
\left(\begin{array}{cc} {1\over 2} & 0\\
                        0 & {1\over 2}\end{array}\right).
                        \end{eqnarray}
                     This third case is more probable, and there is 50\% probability to
occur, whereas case 1 and 2 each has 25\% probability to occur.

Though the three ensembles are distinguishable, they can not be
exploited to realize faster-than-light quantum communication
because one can not ascribe a unique compressed density matrix to
a measurement because of the fluctuation in the quantum
measurement. Impossibility of faster-than-light communication does
not imply the indistinguishability of ensembles having the same
compressed density matrix, which has been discussed
recently.\cite{wangc}

 \section*{5. DISTINCTION OF ENSEMBLES HAVING THE SAME COMPRESSED DENSITY MATRIX}
 \label{s4}

 \subsection*{5.1 Present STATUS}

 \item\label{ss16p} Penrose doubts the completeness of the density
 matrix
description. In an example where A (I myself) on Earth and B (my colleague) on the Moon
each hold a particle from an EPR pair, Penrose states that "However, it is hard to take
the position that the density matrix describes reality. The trouble is that I do not
know that I might not, sometimes later, get a message from the Moon, telling me that my
colleague actually measured the state and found the answer to be such and such. Then, I
know what my particles's state must actually be. The density matrix did not tell me
everything about the state of my particle. For that, I really need to know the actual
state of the combined pair. So, the density matrix is sort of provisional description,
and that is why it is sometimes called FAPP(i.e., for all practical
purposes)."\cite{penrose}

Here Penrose refers to what we call the reduced density matrix for
a composite quantum system of two constituent particles. In our
later discussion, we mainly focus our attention to the compressed
density matrix.

\item  It is commonly believed that ensembles with the same
compressed
 density matrix are physically identical.  We believe that this
 misbelief can largely be eliminated once a clear distinction
 between the compressed and the full density matrices is
 established.

A complexity  arises due to the indistinction of proper mixed
state and improper mixed state.   Here we are talking about the
distinction of ensembles
 with the same compressed density matrix rather than different composite
 systems with the same reduced density matrix. At present, there is intensive
  research work on quantum entanglement based on reduced density matrix for
  composite quantum systems.  The distinction of ensembles having the same compressed
  density matrix we address here is not concerned with coupled quantum systems.

 After clearing up those confusions in the terminology, there are
 still some authors who believe that ensembles with the same
 compressed density matrix are physically identical.

  \item\label{s15} Asher Peres stresses that ensembles with
  infinite number of particles
 having the same compressed density matrix could not be
 distinguished physically. He uses an example to illustrate this
 idea. In one ensemble with infinite number of photons, each photon
 is prepared in either the $|H\ket=|0\ket$ state or $|V\ket=|1\ket$ state
 according to the result of coin-tossing. Hence the compressed
 density of this ensemble is $\rho={1\over 2}{\rm 1_2}$. In another
 ensemble of infinite number of photons, each photon is prepared
 in the left-handed circular or right-hand circular state according to
 the result of coin-tossing, and the compressed density matrix is
 also $\rho={1\over 2}{\rm 1_2}$. Peres argued
 that there are no physical means to distinguish these
 two ensembles. He has elevated this into a
 fundamental postulate in quantum mechanics
 \begin{quote}
 The (compressed ) $\rho$ matrix specifies all the physical properties of a
 quantum ensemble (with infinite number of molecules).
 \end{quote}
 Here the words inside the parenthesis are added by the authors to conform to the
 terminology used in this article.

 At first sight, it may seem that this fundamental postulate excludes
 any possibility of distinguishing ensembles with the same compressed density
 matrix. However, it is not so because Peres' definition of ensemble   is different from
 ours.
 Peres' ensemble contains an infinite number of molecules and
 it only exists conceptually as stressed by Peres\cite{peres}.
 Thus it does not exclude the possibility of distinguishing
 ensembles with a finite number of molecules having  the same compressed density matrix.

 In fact, Peres stated that finite particle number ensembles with
 the same compressed density matrix can be physically
 distinguished \cite{peres2}. In this example, Peres constructed
 two ensembles with 1 million qubits each. In one case, half of
 the qubits are polarized in $|0\ket$ state and the other half in
 state $|1\ket$. In another case, half of the qubits are in state
 ${1\over\sqrt{2}}(|0\ket+i|1\ket)$ and the other half in state
 ${1\over\sqrt{2}}(|0\ket-i|1\ket)$. With high probability, people
 can distinguish the two ensembles (assemblies in Peres's term).
 The probability of failure is proportional to ${1\over \sqrt{N}}$
 where $N$ is the number of molecules in the ensemble. Apparently,
 as $N$ increases, the probability of failure is approaching zero.

 \item Preskill stresses that differently prepared
 ensembles with the same compressed density matrix are not
 distinguishable on one hand, but he also points out different preparations of ensembles
 with the same compressed density
 matrix could be distinguished physically with some additional external
 help\cite{preskill}. It is noted that Preskill's ensemble has a fixed number of molecules.
 Explicitly, Preskill proposed a method to
 distinguish the ensembles $A_2$ and $A_3$ in section 4. He
 presumably assumes that the compressed density matrix of $A_2$ and
 $A_3$ are the same, which is usually not true when the number of
 particles $N$ is a fixed number.
 Suppose Bob starts from ensemble $B_1$  and
 makes a $\sigma_x$  measurement  on his B
 qubit molecules and hence prepares the ensemble $B_3$ and $A_3$.
 Alice can find out what ensemble her ensemble is with some help
 from Bob: Bob tells Alice the result of the measurement for
 each B qubit molecule, whether up or down, but he does not tell
 her what measuring basis he has used. Alice can choose any
 measuring basis, whether $\sigma_x$ or $\sigma_z$, to measure each of
 her A qubit. If Alice chooses $\sigma_x$, her result will
 have perfect accordance with Bob's. If she chooses $\sigma_z$,
 only 50\% of the results agree with Bob's. Hence Alice "does have a way to distinguish Bob's two preparation
methods"\cite{preskill}.

 \item\label{ss16} d'Espagnat pointed out that ensembles
 having the same compressed density matrix can be distinguished
 physically by observing fluctuations of some observables. He has
 illustrated this in an ensemble of $N$ qubits\cite{despagnat}. In
 ensemble $S_1$, a half of qubits are in state $|0\ket$ and
 another half are in state $|1\ket$ where $|0\ket$ and $|1\ket$
 are the eigenstates of operator $\sigma_z$, the Pauli matrix. In
 ensemble $S_2$, a half of qubits are in state $|+x\ket$ and
 another half are in states $|-x\ket$ where $|+x\ket$ and $|-x\ket$
 are the eigenstates of operator $\sigma_x$.
 In both ensembles, the compressed density matrix is $\rho={1\over
 2}{\rm 1_2}$. Although the average value of any observable $\Omega$,
 in our terminology in a sampling measurement,
  for
 both ensembles are all expressed as
 \begin{eqnarray}
 \bra \Omega\ket={\rm Tr}(\rho \Omega),
 \end{eqnarray}
 the fluctuation of  an observable will
be different for ensembles with different compositions. Translated
into our language,  d'Espagnat actually refers to the fluctuation
of a global measurement is different for different ensemble
compositions with the same compressed density matrix. d'Espagnat
used the following observable
 \begin{eqnarray}
 \Sigma_z=\sum_{i=1}^N\sigma_z(i),
 \end{eqnarray}
where the summation over $i$ is made over  all the molecules in the ensemble. For
ensemble $S_1$, the fluctuation of $\Sigma_z$ is zero,
\begin{eqnarray} (\Delta\Sigma_z)_E=0,\end{eqnarray} because each
molecule is in an eigenstate of $\sigma_z$ and every measurement
of $\Sigma_z$ on the whole ensemble will give identical result.
However for ensemble $S_2$, the fluctuation is
\begin{eqnarray}
(\Delta\Sigma_z)_E=\sqrt{N},
\end{eqnarray}
namely if one measures $\Sigma_z$ on the whole ensemble, different
result may be obtained at different times.  By observing the
fluctuation one can distinguish between the two ensembles with the
same compressed density matrix.

\subsection*{5.2 ENSEMBLES with the same FULL DENSITY MATRIX are identical}\label{ss5.1}

 \item\label{s14} Suppose one treats the whole ensemble having $N$
 molecules  as a single quantum system, the ensemble is described by a full density matrix.
 If each
 molecule in the ensemble is independent and non-interacting, then the state of the  ensemble is
 given by \begin{eqnarray}
|\psi_E\ket=|\psi_1(1)\ket|\psi_1(2)\ket\cdots|\psi_1(N_1)\ket|\psi_2(N_1+1)\ket\cdots|
\psi_2(N_1+N_2)\ket\cdots ,\label{estate}\end{eqnarray}
 where the number inside the round bracket is the
labelling of the molecule and the subscript refers to the state in
which the molecule is in. The state $|\psi_E\ket$  is the product
of states of individual $N$
 molecules. The density matrix for the whole ensemble of $N$ molecules is
 $\rho=|\psi_E\ket\bra\psi_E|$.  Since the wave function contains all
 the information about the system, the corresponding density
 matrix completely describes the state of the whole ensemble.
 We  conclude that ensembles with the same full density matrix are
 physically identical. In this case, the
 density matrix is always one representing a pure state of a quantum system with $N$
 molecules. {\bf If the full density matrices of the two
 ensembles are the same, then the two ensembles are physically
 identical.}

\subsection*{5.3 ENSEMBLES having the same COMPRESSED DENSITY MATRIX
but with different PARTICLE NUMBERS are physically distinguishable}\label{ss5.2}

\item\label{ss18} First, let's discuss the case of two ensembles
having the same compressed density matrix but with different
molecule numbers. With ideal measuring sensitivity, one can
measure up to the accuracy of single molecule. Then from the
expectation value of a global measurement alone, one can
distinguish the two ensembles. The expectation of any observable
with the form
\begin{eqnarray}
\Omega=\sum_{i=1}^N\Omega(i),
\end{eqnarray}
will be sufficient to distinguish the two ensembles, because the
expectation of observable $\Omega$ is
\begin{eqnarray}
\bra \Omega\ket_E=N{\rm Tr}(\rho\Omega). \end{eqnarray} Because
different molecule number $N$ will give different expectation
value, one can distinguish these ensembles from the expectation
value directly. This is easy to understand if we compare two
different amount of samples in nuclear magnetic resonance
experiment. In NMR, the measured quantity is the total
magnetization, and is equivalent to the signals in coil.  The
larger the sample, the stronger the measured signal.

\subsection*{5.4 General EXPRESSION for the FLUCTUATION of OBSERVABLES in GLOBAL MEASUREMENT}

\item\label{ss19} Generalizing the idea of d'Espagnat, ensembles
having the same compressed density matrix could be distinguished
by inspecting the fluctuation of some observables in global
measurement. By this method, we can distinguish ensembles upto
different physical composition: the number of molecules $N_i$ in
state $|\psi_i\ket$.

Suppose there are $N$ molecules in an ensemble with $N_k$ molecules in state
$|\psi_k\ket$, and $k=1,\cdots,m$. The fluctuation of the observable
\begin{eqnarray}
\Omega=\sum_{i=1}^N\Omega(i),
\end{eqnarray}
where $i$ is the molecule index in the ensemble, is
\begin{eqnarray}
(\Delta \Omega)_E=\sqrt{N({\rm
Tr}(\Omega^2\rho)-\sum_{k}N_k\bra\psi_k|\Omega|\psi_k\ket^2},\label{etrue}
\end{eqnarray}
where the summation is over the possible single molecule states in the ensemble. It is
apparent that the second term under the square-root is composition sensitive: ensembles
with different compositions  will give different results. This can be comprehended in
the following way.  For a single molecule in state $|\psi_k\ket$, the fluctuation of
$\Omega$ squared is
\begin{eqnarray}
(\Delta\Omega)_k^2=\bra\psi_k|\Omega^2|\psi_k\ket-\bra\psi_k|\Omega|\psi_k\ket^2,\label{measure}
\end{eqnarray}
and contribution from $N_k$ such molecules is $N_k(\Delta\Omega)^2_k$, and the total
contribution from all the molecules in the ensemble is $\sum_k N_k(\Delta\Omega)_k^2$.
With some simple calculation, this gives the result in Eq. (\ref{etrue}).

This result can be derived directly if one recalls that the state
of the whole ensemble is Eq.(\ref{estate})
 The fluctuation of observable $\Omega$, namely the fluctuation of
 the global measurement
is
\begin{eqnarray}
(\Delta\Omega)_E&=&\sqrt{\bra
\Psi_E|\Omega^2-(\bra\Omega\ket)_E^2|\psi_E\ket}.
\end{eqnarray}
The expression inside the square-root can be calculated and it is
\begin{eqnarray}
&&\bra\psi_E|\left(\sum_i \Omega(i)^2+\sum_{i\neq
j}\Omega(i)\Omega(j)-\sum_i\bra\psi_E
|\Omega(i)|\psi_E\ket^2-\sum_{i\neq
j}\bra\psi_E|\Omega(i)\Omega(j)|\psi_E\ket\right)|\psi_E\ket\nonumber\\
&=&\sum_{k}\bra\psi_k|\Omega^2|\psi_k\ket-\sum_{k}
N_k\bra\psi_k|\Omega|\psi_k\ket^2,
\end{eqnarray}
where $i$ runs over the molecules in the ensemble and $k$ runs
over all possible single molecule states in the ensemble.

\subsection*{5.5 Composite QUANTUM SYSTEMS with the same REDUCED DENSITY MATRIX}

\item\label{ss20} Different composite quantum systems may produce
the same reduced density matrix. They could be distinguished if
one measures quantities involving the whole composite quantum
system. For instance let us consider two systems of an EPR pair
$A$ and $B$, one  in  state
\begin{eqnarray}
|\psi^1_{AB}\ket=\sqrt{1\over
2}\left\{|0_A0_B\ket+|1_A1_B\ket\right\},
\end{eqnarray}
and another  in state
\begin{eqnarray}
|\psi^2_{AB}\ket=\sqrt{1\over
2}\left\{|0_A0_B\ket-|1_A1_B\ket\right\}.
\end{eqnarray}
In both cases, the reduced density matrix is $\rho_A={1\over
2}{\rm 1_2}$. By measuring the observables of molecule A alone,
one can not distinguish these two different composite quantum
systems. However, the two systems will be distinguished the moment
one makes a joint Bell-basis measurement.

\section*{6. THE NATURE OF ENSEMBLE NUCLEAR MAGNETIC RESONANCE
QUANTUM COMPUTING}\label{s6}

\item\label{sso21} Ensembles with the same compressed density
matrix can be distinguished physically is very important in
quantum information processing. One important issue is the nature
of room-temperature NMR quantum computing. It has been widely
accepted that NMR quantum computing is not quantum mechanical
because there is no entanglement in each step of the NMR quantum
computing\cite{braunstein}. We point out here that this conclusion
is based solely on the unjustified belief that ensembles having
the same compressed density matrix are identical.

The major argument in Ref.\cite{braunstein} is to rewrite the effective density matrix,
which is the compressed density matrix of the effective pure state obtained after the
effective pure state technique,
\begin{eqnarray}
\rho=\left({1-\epsilon \over 4}\right) I_d+\epsilon\rho_{eff},
\end{eqnarray}
where $\rho_{eff}$ is the effective pure state density matrix, and
$\epsilon$ is a number about $10^{-6}$ large and $I_d$ is $d\times
d$ unit matrix, into a convex decomposition of product-state
density matrices,
\begin{eqnarray}
\rho=\sum_{i}C_i\left(\rho_1\bigotimes\cdots\rho_M\right)^i,
\label{efffrho}
\end{eqnarray}
where $C_i$ is the decomposition coefficient and
$\left(\rho_1\bigotimes\cdots\rho_M\right)^i$ is the density
matrix representing the product of $M$ qubits. This convex
decomposition is possible when $\epsilon$ is small and the qubit
number $M$ is small. Since the product state does not have
entanglement, then it was concluded by the equivalence of
ensembles having the same compressed density matrix that the
current NMR quantum computing does not have quantum entanglement,
and hence is not genuine quantum mechanical.

From the previous discussions in this article, we see that the
conclusion is based on an unjustified belief. Though the
compressed density matrix of the effective pure state could be
rewritten into convex decomposition of products state density
matrices, they are not physically equivalent. As a simple example,
we show that the effective Bell-state ensemble and its product
state decomposition in Ref.\cite{braunstein} can be distinguished
physically. The effective Bell-state compressed density matrix is
\begin{eqnarray}
\rho={1-\epsilon \over 4}I_4+\epsilon\rho_{Bell},\label{rhobell}
\end{eqnarray}
where $\rho_{bell}=(|00\ket+|11\ket)(\bra 00|+\bra 11|)/2$. The
effective Bell-state describes an ensemble with $\epsilon N$
molecules in Bell-state $(|00\ket+|11\ket)/\sqrt{2}$ and
$(1-\epsilon)N/4$ molecules in each of the calculating-basis
states $|00\ket$, $|01\ket$, $|10\ket$ and $|11\ket$. According to
Ref.\cite{braunstein}, it can be decomposed into
\begin{eqnarray}
\rho=\sum_{i,j}{1\over 4}({1\over 9}+C_{ij})P_i\otimes
P_j,\label{rhobell2}
\end{eqnarray}
where $P_i=(1+\sigma_i)/2$ for $i=1,2,3$ which represents a pure
state polarized along the three axis $x$, $y$ and $z$
respectively, and $P_i=(1-\sigma_i)/2$   for $i=4,5,6$ which
represents a pure state anti-polarized along the three axis $x$,
$y$ and $z$ respectively. The coefficient $C_{ij}$ are given in
Table \ref{t1}.
\begin{table}
\begin{center}
\begin{tabular}{crrrrrr} \hline
       & $P_1$ & $P_2$ & $P_3$ & $P_4$ & $P_5$ & $P_6$\\ \hline
 $P_1$ & $\epsilon$ &            &  $-\epsilon$ & & &   \\
 $P_2$ &            & $-\epsilon$ &             & &$\epsilon$ & \\
 $P_3$ &            &             & $\epsilon$  &  &
 &$\epsilon$\\
 $P_4$ &$-\epsilon$ &   &   &$\epsilon$ & & \\
 $P_5$ &            &$\epsilon$ & & &$-\epsilon$ & \\
 $P_6$ &            &           &$-\epsilon$ & & & $\epsilon$\\
 \hline
 \end{tabular}
 \end{center}
 \caption{$C_{ij}$ for the effective pure Bell-state. A space
 means that coefficient is zero.}\label{t1}
 \end{table}
 The two different compositions of the effective Bell-state can be
 distinguished by observing the fluctuation of the operator
 \begin{eqnarray}
 \Sigma_{zz}=\sum_{i}\sigma_{1z}(i)\sigma_{2z}(i).
 \label{ezz}
 \end{eqnarray}
For the ensemble implied by (\ref{rhobell}), the fluctuation is
$(\Delta\Sigma_{zz})_{1,E}=\epsilon\sqrt{ N}$, whereas for the
product state expansion (\ref{rhobell2}), the fluctuation is
$(\Delta\Sigma_{zz})_{2,E}={2\sqrt{N} \over 3}$.

Hence the two ensembles are distinguished though they all have the same compressed
density matrix. The decomposition of the effective Bell-state compressed density matrix
into a product state compressed density matrix can not be used to infer the conclusion
that no entanglement exists in the effective Bell-state. Entanglement is the property
of the individual molecules, not the ensemble as a whole.  As in Ref.\cite{long}, even
if all the molecules in an ensemble are in the same quantum state, there is no
entanglement between different molecules. Hence the effective pure state should be
considered as what it is physically, and there are indeed quantum entanglement within
each molecule, though only a small portion of the molecule from the effective pure
state contribute to the final read-out in the NMR  detection signal.

\item In addition, there is an unnaturalness difficulty in the product state
decomposition description of the NMR quantum computing used in Ref.\cite{braunstein}.
In the effective pure state NMR quantum computing, the quantum computation is performed
on the $\epsilon N$ molecules in the effective pure state. In the product state
decomposition, if only single qubit operations are performed one can still retain a
simple picture. We can visualize that there are several "tubes" in the NMR sample. For
instance in a two-qubit system case, there are 36 "tubes" in the sample, each tube
contains some number of molecules in a product state $P_i\otimes P_j$. A single qubit
operation just changes the single molecule state into another state so that one can
still use the same "tubes" to describe the state of the ensemble.  However once an
entangling operation such as the controlled-NOT gate is performed, the number of
molecules within each composition, that is in a "tube",  will change. For example, if
we perform a CNOT gate on effective Bell-state (\ref{rhobell}), this changes the
ensemble into an effective $|+x,0\ket$-state, that is, the first qubit is in $|+x\ket$
state and the second qubit is in $|0\ket$ state,
\begin{eqnarray}
\rho_{x}={(1-\epsilon) \over 4}I_4+\epsilon|+x,0\ket\bra
+x,0|,\label{rhox}
\end{eqnarray}
which is natural and clear. However in the product state expansion
of Ref.\cite{braunstein}, this density matrix becomes
\begin{eqnarray}
\rho_{x}'=\sum_{i,j}{1\over 4}({1\over 9}+C'_{i,j})P_i\times
P_j,\label{rhox2}
\end{eqnarray}
where the coefficients $C'_{i,j}$ are given in Table \ref{t2}.
\begin{table}
\begin{center}
\begin{tabular}{crrrrrr} \hline
       & $P_1$ & $P_2$ & $P_3$ & $P_4$ & $P_5$ & $P_6$\\ \hline
 $P_1$ & $\epsilon/3$ & $\epsilon/3$ & $5\epsilon/3$ &$-\epsilon/3$ & &   \\
 $P_2$ &            &  & $\epsilon/3$ & & & \\
 $P_3$ &            &             & $\epsilon$  &  & &\\
 $P_4$ &$-\epsilon/2$ &$-\epsilon/3$   &$-\epsilon$   &$-\epsilon/3$ & $-\epsilon/3$&$\epsilon$ \\
 $P_5$ &            &  & $\epsilon/3$& & &$-\epsilon/3$ \\
 $P_6$ &            &           &$\epsilon/3$ & & & $-\epsilon/3$\\
 \hline
 \end{tabular}
 \end{center}
 \caption{$C'_{ij}$ for the effective pure state $|+x,0\ket$. A space
 means that coefficient is zero.}\label{t2}
 \end{table}
 We see that there is a reshuffling of the number of molecules in
 different states. In order to get the same averaged result of the effective $|+x,0\ket$ state,
 one has to reshuffle the distribution of the molecule numbers in different "tubes".
  While for non-entangling operations, there is
 no need for such a reshuffling because the single-qubit operation
 can be performed on each composition of states directly, and
 their effect is to change state $P_i$ to state
 $P'_i=UP_iU^{\dagger}$.

\item One objection to Eq. (\ref{measure}) is that each molecule
in an ensemble is in a  mixed state itself, even at a given
instant. In this case, the fluctuation of observable $\Omega$ is
\begin{eqnarray}
\Delta\Omega_E=\sqrt{N{\rm Tr}(\rho\Omega^2)-N({\rm
Tr}(\rho\Omega)^2}.
\end{eqnarray}
In this case, all molecules in the ensemble are equivalent and no
physical means can distinguish them. In fact, there is no
difference at all, because the compositions of the ensembles are
the same.  This has been used by some authors to criticize
d'Espagnat \cite{tolar}. To this objection, we stress that it is
impossible to prepare such an ensemble using classical
probabilities. For instance, in the example given by Peres
\cite{peres}, the state of a photon is prepared according to the
result of a coin-tossing, either along $z$ or along $-z$. It is
true that each photon has 50\% probability to be prepared in state
$|0\ket$ or $|1\ket$. But each photon is prepared in a definite
state, not in a  mixed state. This is exactly the case  in the
BB84 quantum key distribution protocol\cite{bb84} where the state
of a qubit has 25\% probability in each of the four states
$|z\ket$, $|-z\ket$, $|x\ket$ and $|-x\ket$. But it is certainly
in one of the four possible states.

One possible way to realize such a scenario is that each molecule is in an improper
mixed state. For instance in an ensemble of molecules, each molecule contains two
qubits $A$ and $B$, and the two qubits are in an entangled state, say
\begin{eqnarray}
|\psi_{AB}\ket={1\over
\sqrt{2}}\left(|00\ket_{AB}+|11\ket_{AB}\right).
\end{eqnarray}
If one looks at the state of the first qubit in each molecule in
the ensemble, the resulting state is  an improper mixed state.
Then in this case, by observing the fluctuations of observables
for qubit $A$ alone one can not distinguish another ensemble with
the same compressed density matrix where $A$ and $B$ are in state
\begin{eqnarray}
|\psi'_{AB}\ket={1\over
\sqrt{2}}\left(|01\ket_{AB}+|10\ket_{AB}\right).
\end{eqnarray}

However, the above picture is not true in NMR quantum computing.
Gershenfeld and Chuang have explicitly explained that in effective
pure state there are some numbers of molecules in a definite
quantum state\cite{gersh}. This is a valid approximation for NMR
 quantum computing because nuclear spins are well isolated from the environment,
hence it has a long decoherence time. For those nuclear spins that
are not used in quantum computing, decoupling pulses are usually
used to disentangle them from the working nuclear spins so that
they can not form entangled state that will make the working
qubits in an improper mixed state. Thus a valid picture of the
ensemble used in NMR quantum computing is that the ensemble
contains lots of molecules, there are certain number of molecules
in each of the quantum states of the molecule. The spin-relaxation
time $T_1$ can be viewed as the time period a molecule remains at
a given quantum state. This is supported by he Gorter formula
\cite{gorter}
\begin{eqnarray}
{1\over T_1}={1\over 2} {\sum_{n,m}W_{n,m}(E_m-E_n)^2 \over \sum_n
E_n^2},
\end{eqnarray}
where $W_{m,n}$ is the transition probability rate from level $m$
to level $n$. Hence $T_1$ is the energy weighted time a molecule
remain at a given quantum state.

What is then the significance of studying quantum entanglement
directly from the density matrix? The study of entanglement for
the compressed density matrix directly is  of significance because
for those compressed density matrices that can not be decomposed
into product states, their microscopic entanglement are already
reflected in the expectation value.  However for ensembles, it is
more important to study the entanglement of the pure state of
single molecules because it is the place where quantum
entanglement exists.  As we see, even though the expectation value
does not show the entanglement, one can still exploit this
precious resource through some method. On the other hand, for
improper mixed state, the study of entanglement based on density
matrix is of  importance.

\section*{7. SUPPORTING EXAMPLE FROM QUANTUM COMMUNICATION}

\item Though fluctuations of observables in ensembles with
infinite number of photons can not distinguish different
preparations, there are still methods to distinguish them if one
has individual access to the photons in the ensemble. The quantum
key distribution is  a good example. First we briefly describe the
BB84 quantum  key distribution scheme\cite{bb84}. In the BB84
scheme,  Alice sends a sequence of photons to Bob. Each photon is
randomly in one of the four states: $|0\ket$, $|1\ket$, $|+x\ket$,
$|-x\ket$ respectively. Hence the compressed density matrix of the
ensemble, with infinite number of photons theoretically, is
$\rho={1\over 2}{\rm 1_2}$. Then Bob chooses randomly one of the
two measuring devices, the $\sigma_z$-basis or $\sigma_x$-basis to
measure the state of the photons. After the transmission, Alice
and Bob publicly compare their measuring-basis used for each
photon, and they retain those events in which they choose the same
measuring-basis. From these retained events, they choose a
sufficiently large subset of events and publicly compare the
measuring results of these events. If the error rate is high, they
conclude that there is Eve during the transmission and discard the
results. Otherwise they proceed to the post-processing stage and
get a final secret key.

If the compressed density matrix had described all the physical
properties of the ensemble, we would run into difficulty in the
security of quantum key distribution. The same compressed density
matrix could also be produced by Alice with only two states
$|0\ket$ and $|1\ket$: she randomly prepares the photons in one of
these two states. In this way, the compressed density matrix is
the same $\rho={1\over 2}{\rm 1_2}$.  Immediately we know that
this replacement is invalid: the simplified scheme is unsafe, for
if Eve uses the $\sigma_z$ measuring device to measure every
photon, she can steal every bit of information. Then  she can
resends a photon with the same state to Bob. In this way, her
action could not at all be detected. Hence the scheme becomes
insecure. This conclusion would not alter if we had used infinite
number of photons in the ensemble.

\section*{8. SUMMARY}

\item We have examined in details the definition of ensembles in quantum mechanics and
the different definitions of mixed states. We have proposed to use different
terminologies for the density matrix in different situations: the full density matrix,
the compressed density matrix and the reduced density matrix. We have explicitly shown
that ensembles having the same compressed density matrix but different compositions can
be distinguished physically by observing fluctuations of observables in the whole
ensemble. With this conclusion at our hand, we studied the problem of the nature of NMR
quantum computing. We have pointed out that the conclusion that there is no
entanglement in the current NMR quantum computing experiment is based on an unjustified
belief that ensembles having the same compressed density matrix are physically
equivalent. This conclusion is also supported by the security aspect of quantum key
distribution where the density matrix $\rho$ could not specify all the physical
properties of an ensemble, even if the number of molecules in the ensemble is infinite.
A brief summary is given in  Table \ref{t3}.

 GLL are grateful to  Professors C. N. Yang,  Alain Aspect
, Peter Zoller, Hans Briegel, Li You, C P Sun, K. N. Huang, Brian J Dalton for helpful
discussions. They also thank X. B. Wang, Minki Jeong, Piero Mana for email discussions.
HWL is supported by a grant from the Korea Science and Engineering Foundation (KOSEF)
through Korea-China International Cooperative Research Program. This work is supported
by the National Fundamental Research Program Grant No. 001CB309308, China National
Natural Science Foundation Grant No. 10325521, 60433050,the Hang-Tian Science Fund, and
the SRFDP program of Education Ministry of China.

\bigskip

\end{enumerate}

\begin{table}
\begin{center}
\caption{A summary}\label{t3}
\begin{tabular}{|p{4cm}|p{12cm}|}\hline
 \multicolumn{2}{ | c|}{     Full density matrix}\\ \hline
Object described & A single quantum system, but may contain many
molecules\\
Matrix Name  & Full density matrix $\rho_f=|\psi\ket\bra\psi|$\\
Object Name  & A pure state. The object is always described by a single wave function,
or
equivalently a full density matrix for a pure state \\
Remark & The description is complete. Different preparations with the same full density
matrix are physically identical.\\ \hline \multicolumn{2}{|c|}{Compressed density
matrix}\\ \hline Object described & An ensemble of $N$ independent molecules with $N_i$
molecules in
 state $|\psi_i\ket$.\\
 Matrix Name  & Compressed density matrix
 $\rho_c=\sum_i{N_i\over N}|\psi_i\ket\bra\psi_i|$. $\rho_c$ describes the state of an averaged molecule from this
 ensemble. The full density of the whole ensemble is $\rho_f=\rho_1\otimes\rho_2\otimes\cdots\otimes\rho_N$,
 where the state of different molecules can be the same. Though two molecules may be in the same quantum state,
 they are distinguishable, for instance by their positions in space. The relation between the
 full density matrix and the compressed density matrix is  $\rho_c={1\over N}\sum_i \rho_i$, where $\rho_i$ is the full
 density matrix for molecule $i$ in the ensemble. \\
 Object Name & The object described by $\rho_c$ is called a proper
 mixture. \\
Remarks & One can not tell the difference between two ensembles having the same
compressed density matrix  by sampling measurement or by the average value,
 but they can be distinguished by observing the
 fluctuations of observables regarding the whole ensemble.  \\ \hline
\multicolumn{2}{|c|}{Reduced density matrix}\\ \hline Object
described & Composite quantum systems \\
Matrix Name & Reduced density matrix which is
 obtained by taking trace over other degrees of freedom from the
 full density matrix of the whole composite system. \\
Object Name & improper
 mixture. \\
 Remark & By observing the observables of only a part from the
 composite system, it is impossible to distinguish systems with
 the same reduced density matrix. But the composite systems can be
 distinguished if the whole composite systems are measured.\\
 \hline
 \end{tabular}
 \end{center}
 \end{table}

\end{document}